\documentclass[epj]{webofc}
\usepackage[varg]{txfonts}   
\wocname{EPJ Web of Conferences}
\woctitle{CONF12}
\newcommand{\Det}{{\rm Det}}
\newcommand{\bea}{\begin{eqnarray}}
\newcommand{\eea}{\end{eqnarray}}
\newcommand{\Tr}{{\rm Tr}}
\newcommand{\tr}{{\rm tr}}

\begin{document}
\selectlanguage{english}
\title{Heavy-heavy and heavy-light quarks interactions generated
  by QCD vacuum}

\author{Mirzayusuf Musakhanov\inst{1}\fnsep\thanks{\email{musakhanov@gmail.com}}}

\institute{National University of Uzbekistan}

\abstract{
 The QCD vacuum is populated by instantons that correspond to the 
tunneling processes in the vacuum. This mechanism creates the strong
vacuum gluon fields. As result, the QCD vacuum instantons induce very
strong interactions between light quarks, initially almost
massless. Such a strong interactions bring a large
dynamical mass $M$ of the light quarks and bound them to  
produce almost massless pions in accordance with the spontaneous
breaking of the chiral symmetry (SBCS). On the other hand, the QCD
vacuum instantons also interact with heavy quarks and responsible for
the  generation of the heavy-heavy and heavy-light quarks
interactions, with a traces of the SBCS. 
If we take the average instanton size $\bar\rho=0.33$ fm, 
and the average inter-instanton distance $\bar R=1$ fm  
we obtain the dynamical light quark mass to be $M = 365$ MeV and
the instanton media contribution to the heavy 
quark mass $\Delta M$=70 MeV. These factors define the coupling
between heavy-light and heavy-heavy quarks induced by the QCD vacuum 
instantons.  

We consider first the instanton effects on the heavy-heavy quarks potential,
including its spin-dependent part. We also discuss those effects
on the masses of the charmonia and their hyperfine mass splittings.  
At the second part we discuss the interaction
between a heavy and light quarks generated by instantons and it's effects.

}
\maketitle
\section{Introduction}
\label{intro}
QCD vacuum has a rich topological properties. We know from~\cite{FJR1976} that the 
classical potential energy of the gluon field $A$ is periodic along their collective coordinate so-called Chern-Simons number $N_{CS}=\int\!\! d^3x K_4 $ (where topological current 
$ K_\mu\!=\!\frac{1}{16\pi^2}\epsilon_{\mu\alpha\beta\gamma}
\left(\!A_\alpha^a\partial_\beta A_\gamma^a\!\!+\!\!\frac{1}{3}\epsilon^{abc}
A_\alpha^a A_\beta^b A_\gamma^c\!\right)\!\,).$
and oscillator-like in all other directions in functional space.
Quasi-classical quantization leads to the band structure. The width of the band is defined by 
the amplitude of tunneling $\sim e^{-S_I}$, where the action 
$S_I$ is calculated on Euclidean classical trajectory between nearest minima -- QCD (anti)instanton. 
It means that (anti)instanton's topological charge  $
Q_T=\frac{1}{32\pi^2}\int d^4x\; F_{\mu\nu}^a {\tilde F}_{\mu\nu}^a
=N_{CS}(+\infty)-N_{CS}(-\infty)=\pm 1$. Here $ F_{\alpha\beta}\equiv \frac{1}{ig}[iD_\alpha,iD_\beta],\,\,\, iD_\alpha\equiv i\partial_\alpha+gA_\alpha, \,\,\,
{\tilde F}_{\mu\nu}^a \equiv \frac{1}{2}\epsilon_{\mu\nu\alpha\beta}F_{\alpha\beta}^a.$
Then, we may assume that in QCD vacuum 4-dim space is filled by a tunneling processes with the density $n=1/R^4$ and 
time-space size $\rho$. It correspond to the collection of a (anti)instantons with sizes $\rho$ occupied Euclidean space 
with interinstanton distances $R$ and total vacuum gluon field is given by 
$A_{\mu}(x)=\sum_{+}^{N_{+}}A_{\mu}^{I}(\zeta_{+},x)+\sum_{-}^{N_{-}}A_{\mu}^{A}(\zeta_{-},x)$, 
where $A_{\mu}^{I}(\zeta_{+},x)$ ($A_{\mu}^{A}(\zeta_{-},x)$) is (anti)instanton -- BPST selfdual classical solution of the equations of motion in Euclidean space~\cite{BPST1975}. Its collective coordinates are $\zeta=(\rho,z,U)$ -- size, position and color orientation respectively.

In general, $A_{\mu}(x)$ is a not a solution of the equations of motion and it leads to the interactions between instantons.
The quantization on this background also is another source of the inter-instanton interaction. The main assumption of the Instanton Liquid Model (ILM) that these factors stabilize the $\rho$ and $R$.  
The estimates of the averaged instanton size $\bar \rho$;
and averaged inter-instanton distance $\bar R$ show 
\begin{eqnarray}
 &  & \bar\rho\simeq0.33\, \mathrm{fm},\, \bar R\simeq1\,
      \mathrm{fm},
      \mbox{(phenomenological)}~~
      \mbox{\cite{Diakonov:2002fq,Schafer:1996wv}}, 
\label{classicalParameters}\\    
 &  & \bar\rho\simeq0.35\, \mathrm{fm},\, \bar R\simeq0.95\,
      \mathrm{fm},\mbox{(variational)}~~\mbox{\cite{Diakonov:2002fq}},
\nonumber   \\ 
 &  & \bar\rho\simeq0.36\, \mathrm{fm},\, 
 \bar R\simeq0.89\, \mathrm{fm},~\mbox{(lattice)}~\mbox{\cite{lattice}},
\nonumber   \\ 
 &  & \bar\rho\simeq0.35\, \mathrm{fm},\, 
 \bar R\simeq0.86\, \mathrm{fm},~\mbox{(ILM with $1/N_c$ corrections)}~\mbox{\cite{Goeke:2007bj}}.
  \nonumber \end{eqnarray}
They coincide within $10-15\%$ uncertainty.
With packing parameter $\pi^2 (\frac{\bar \rho}{\bar R})^4 \sim 0.1-0.3$ we have a comfortable possibility 
of independent averaging over instanton positions and orientations.

\begin{figure}[h]
\centerline
{
\includegraphics[scale=0.3]{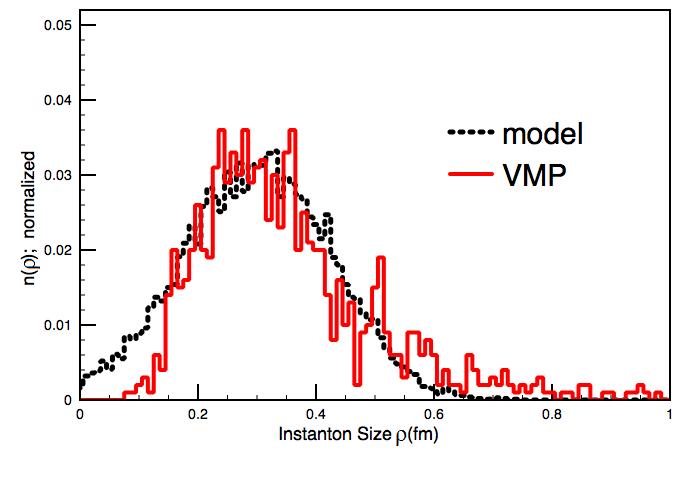}
}
\vskip -0.3cm
\caption
{Instanton size distribution from lattice vs ILM~\cite{instantonsize}.}
\label{instantonsize}     
\end{figure}
\vskip -0.6cm
 It is clear that in ILM it was neglected by possible portion of large instantons providing confinement (see Fig.~\ref{instantonsize}).
The progress with understanding of the confinement is related with extension of a BPST instantons to a KvBLL instantons, which are described   by superposition of a dyons~\cite{KvBLL1998}. Then, the ILM is extended to the  Dyon Liquid Model (DLM)~\cite{Diakonov2009,Shuryak2015}. 
 At SU(2) there is two types of dyons. At large separations $r$ the action density
is a clear superposition of these dyons, while at small separations they merge and become like instanton. 
The KvBLL instanton size and the dyons separation are related by $\pi T \rho^2=r$, where $T$ is temperature. 
The estimate within  DLM leads the
KvBLL instanton average size $\bar\rho\approx \frac{\alpha_s N_c}{2\pi\Lambda_{PV} }\approx 0.5\,{\rm fm}$ at $N_c=3$, $\alpha_s = 0.5$, $\Lambda_{PV}=200\,{\rm MeV}$~\cite{Diakonov2009}. 

On the other hand, quark core sizes of lowest states of light and heavy quarks hadrons are small. 
The estimates of the sizes of  lowest states of a heavy quarkonium within non-relativistic potential model~\cite{quarkoniumsize} 
are demonstrated at Table~\ref{tablesizes}.
\begin{table}[h]
\centering\scalebox{0.9}{
\begin{tabular}{|c|c|c|c|c|c|c|}
\hline 
State & $J/\psi$ & $\chi_c$ & $\psi'$ & $\Upsilon$ & $\chi_b$ & $\Upsilon'$
\tabularnewline
\hline  
mass [Gev] & 3.07 & 3.53 & 3.68 & 9.46 & 9.99 & 10.02
\tabularnewline
\hline 
size $r$ [fm] & 0.25 & 0.36 & 0.45 & 0.14 & 0.22 & 0.28
\tabularnewline
\hline
\end{tabular}
}
\label{tablesizes}
\caption{Quarkonium states and its sizes in non-relativistic potential model~\cite{quarkoniumsize}.} 
\end{table}  

The same conclusion is on the nucleon.
A model estimates from various nucleon form-factors data show nucleon quark core size  $r_N\sim 0.3-0.5$ fm~\cite{nucleon-quarkcore}. 

Since, small quark core size hadrons are insensitive to the confinement,
we may safely apply ILM.

\subsection{Light quarks physics}
We assume that  light quark current mass $m$ is small. It means $m\bar\rho<<1.$ Than zero-mode
$|\Phi_{0i}>$, obtained from the Dirac Eq. $(\hat p+g\hat A)|\Phi_{0i}>=0$, become dominant component
in single instanton field light quark propagator 
$S_{i}$. Accordingly the
interpolation formula~\cite{Musakhanov,Goeke:2007bj}: 
\begin{eqnarray}\label{Si}
S_{i}=S_{0}+S_{0}\hat{p}\frac{|\Phi_{0i}><\Phi_{0i}|}{c_{i}}\hat{p}S_{0},\,\,\, S_{0}=\frac{1}{\hat{p}+im},\,\,\,
c_{i}=im<\Phi_{0i}|\hat{p}S_{0}|\Phi_{0i}> .
\end{eqnarray}
 The advantage of this interpolation is shown by the projection of
$S_{i}$ to the zero-modes: 
\begin{eqnarray}
S_{i}|\Phi_{0i}>=\frac{1}{im}|\Phi_{0i}>,\,\,\,<\Phi_{0i}|S_{i}=<\Phi_{0i}|\frac{1}{im}
\end{eqnarray}
as it must be, while the similar projection of $S_{i}$ given by
Ref.~\cite{Diakonov:1995qy} has a wrong component, negligible only in 
the $m\rightarrow0$ limit. This interpolation well reproduce exact solution~\cite{exactSi}.

Summing the re-scattering series we arrive to the total ILM light quark propagator:  
\begin{eqnarray}
 S =S_{0} - S_{0}\sum_{i,j}\hat p |\Phi_{0i}\rangle
\left\langle\Phi_{0i}\left|\left(\frac{1}{B(m)}\right)\right|\Phi_{0j}\right\rangle
\langle\Phi_{0j}|\hat p S_{0},
\,\,
 B(m)= \hat p  S_{0}\hat p.
\label{lightpropagator}
\end{eqnarray}
Our aim is to get light quark determinant   represented as 
$\Det(\hat p+g\hat A+im)=\Det_{{\rm high}}\cdot\Det_{{\rm low}}$, where $\Det_{{\rm high}}$
receive a contribution from fermion modes with Dirac eigenvalues from
the interval $M_{1}$ to the Pauli--Villars mass $M_{PV}$, and $\Det_{{\rm low}}$
is accounted eigenvalues less than $M_{1}$. The product of these
determinants is independent  of the scale
$M_{1}$. However, we may calculate both of them only approximately. The quality of the approach
 is given by the dependence of the
product on $M_{1}$, which serves a check of the approximations ~\cite{Diakonov:1995qy}.

Starting from Eq.~(\ref{lightpropagator})  we were able~\cite{Musakhanov} to find low-frequencies part of light quark determinant in the presence of the light quark sources $(\xi,\xi\dagger)$, which is $\Det_{{\rm low}}\exp{(-\xi^{\dagger}S\xi)}=$
 \begin{eqnarray}
 &&=\int\prod_{f}D\psi_{f}D\psi_{f}^{\dagger} \exp\int\left(\psi_{f}^{\dagger}(\hat{p}\,+\, im_{f})\psi_{f}+\psi_{f}^{\dagger}\xi_{f}+\xi_{f}^{\dagger}\psi_{f}\right)
\prod_{f}\left\{ \prod_{\pm}^{N_{\pm}}V_{\pm,f}[\psi^{\dagger},\psi]\right\} \;,
\label{Det}
\\\label{V}
&&V_{\pm,f}[\psi^{\dagger},\psi]= i\int dx\left(\psi_{f}^{\dagger}(x)\,\hat{p}\Phi_{\pm,0}(x;\zeta_{\pm})\right)\int dy\left(\Phi_{\pm,0}^{\dagger}(y;\zeta_{\pm})\hat{p}\,\psi_{f}(y)\right).
\end{eqnarray}
Fermionic fields $\psi^{\dagger},\psi$ has a meaning of constituent quarks.
Averaging of Eq.~(\ref{Det}) over instantons collective coordinates leads to the light quarks partition function:
 \begin{eqnarray}
 \label{partition}
Z[\xi_{f},\xi_{f}^{\dagger}] =\int\prod_{f}D\psi_{f}D\psi_{f}^{\dagger}\exp{\int\left(\psi_{f}^{\dagger}(\hat{p}+ im_{f})\psi_{f}+\psi_{f}^{\dagger}\xi_{f}+\xi_{f}^{\dagger}\psi_{f}\right)}
\prod_\pm\left(\overline{\prod_{f}V_{\pm,f}[\psi^{\dagger},\psi]}\right)^{N_\pm}
 \end{eqnarray} 
Small packing parameter provided here independent averaging. The quantity
$$\overline{\prod_{f}V_{\pm,f}[\psi^{\dagger},\psi]}=\int d\zeta_{\pm}\prod_{f}V_{\pm,f}[\psi^{\dagger},\psi]$$
 is a non-local t'Hooft-like vertex with $2N_f$-legs, where nonlocality range $\sim\bar\rho.$

 at $N_f=1$ and $N_\pm=N/2$ the Eq.~(\ref{partition}) leads to 
\bea
&&Z[\xi,\xi^{\dagger}]=e^{-\xi^{\dagger}\left(\hat p \,+\, i(m+M(p))\right)^{-1}\xi}
\exp\left[\Tr\ln\left(\hat p \,+\, i(m+M(p))\right)+N\ln\frac{N/2}{\lambda}-N\right],
\label{Z}
\\
&&N=\Tr\frac{iM(p)}{\hat p \,+\, i(m+M(p))},\,\,\, M(p)=\frac{\lambda}{N_c}(2\pi\rho F(p))^2.
\label{M}
\eea
Here the form-factor $F(p) $ 
is given by Fourier-transform of the
zero-mode. The coupling  $\lambda$ and the dynamical quark mass $M(p)$
are defined by  the Eq.~(\ref{M}), where $\Tr(...)=\tr_{c,f,\gamma}\int d^4x <x|(...)|x>$. 

 At $N_f >1$ and in the saddle-point approximation (no meson loops
 contribution) $Z[\xi_f,\xi_f^{\dagger}]$ has a similar to 
 the Eq.~(\ref{Z}) form.   
 
In the lowest order on $1/N_c$ and at $\bar\rho= 0.33$ fm and $\bar R= 1$ fm  $M(0)\approx 365$ MeV, while
at $\bar\rho=0.36\,\mathrm{fm}, \bar R=0.89\,\mathrm{fm}$  $M(0)\approx 570\,\mathrm{MeV}.$
Light quark dynamical mass $M$ can be considered as a strength of a light quark-instanton interaction.

Most important features of  light hadron physics -- Spontaneous Breaking of the Chiral Symmetry (SBCS)
  and Chiral Perturbation Theory (ChPT) is very well reproduced by the Eqs.~(\ref{partition}-\ref{M})~\cite{Goeke:2007bj}. 

\subsection{Heavy quarks in ILM}
 
The heavy quark Lagrangian is given by 
$
L_H=\Psi^{\dagger}(\hat P+im_Q)\Psi,
$ 
where
$P=p+gA$.
We make a Foldy-Wouthuysen transformation
accordingly\cite{Chernyshev:1995gj}: 
$
\Psi(x)=\exp(-m_Q\gamma_4 x_4+O(1/m_Q))Q(x),
$
which leads to 
\bea
L_H=Q^{\dagger}\gamma_4 P_4 Q +Q^{\dagger} Q_1 Q,\,\,\, 
Q_1=\frac{{\vec P}^2}{2m_Q}-\frac{\vec\sigma\vec B}{2m_Q},\,\,\, \vec
B=rot \vec A .  
\eea
Then, (infinitely) heavy quark propagator (Wilson line) in ILM defined as
\bea
w=\int D\zeta\frac{1}{\theta^{-1}-\sum_{i}a_{i}},
\label{w}\eea
where 
$$
a_{i}(t)=iA_{i,\mu}(x(t))\frac{d}{dt}x_{\mu}(t), w_{\pm}=\frac{1}{\theta^{-1}-a_{\pm}},\,\,   <t|\theta|t'>=\theta(t-t').
$$
Pobylitsa~\cite{Pobylitsa1989} derived the Eq. for the quark correlators in instanton vacuum. 
It's application~\cite{DPP1989} to the  $w$ leads to:
\bea
w^{-1}=\theta^{-1} + \sum_i\int d\zeta_i (w-a_i^{-1})^{-1}.
\label{pobylitsa}\eea
In the lowest order on density the  
Pobylitsa Eq. has a solution 
\bea
w^{-1}=\theta^{-1}-\frac{N}{2}\sum_{\pm}\theta^{-1}(w_{\pm}-\theta)\theta^{-1}+O(N^{2}/V^{2}).
\label{lowest}\eea
In fact the effective dimensionless parameter of expansion here is $(\bar\rho/\bar R)^4$, which is comfortably small.

Instanton media contribution to the heavy quark mass is
$$\Delta m_Q=16\pi i_0(0)(\bar\rho^4/\bar R^4)\bar\rho^{-1}/N_c,\,\,\,i_0(0)=0.55.$$ 
At $\bar\rho=0.33\,\mathrm{fm}, \bar R=1\,\mathrm{fm}$ $\Delta m_Q\approx70\,\mathrm{MeV}$ and at
 $\bar\rho=0.36\,\mathrm{fm}, \bar R=0.89\,\mathrm{fm}$  $\Delta m_Q\approx140\,\mathrm{MeV}.$
 
 $\Delta m_Q$ can be considered as a strength of a heavy quark-instanton interaction.

\section{$Q\bar Q$ potential in ILM}
\label{QQpotential}
Application of  an Eq.
similar to the Eq.~(\ref{pobylitsa}) to the Wilson loop in  the lowest order on instanton density
leads to the
static central potential~\cite{DPP1989} for the $Q\bar Q$ in colorless state  
\bea
V_C(r)= \frac{N}{2VN_c}\sum_\pm \int d^3 z_\pm \tr_c 
\left[1-
       P\exp\left(i\int_{L_1} dx_4 A_{\pm,4}^{(1)} \right)
        P\exp\left(-i\int_{L_2}  dx_4 A_{\pm,4}^{(2)}\right)
       \right]_{z_{\pm,4}=0} ,
\label{static}\eea
where $L_{1,2}$ are straight lines parallel to the $x_4$ and separated by distance $r$. $A^{(1,2)}$ are the gauge fields projected onto the lines $L_{1,2}$.

Also, from  $1/m_Q^2$ expansion of the heavy quark propagator~\cite{Eichten1980} it is easy to find
spin-dependent parts of the total potential~\cite{YKTMH}
\bea
V(\vec r) = V_C(r)+V_{SS}(r)
(\vec S_Q\!\cdot\!\vec S_{\bar Q})
+V_{LS}(r)(\vec L\cdot\vec S)
 +V_{T}(r)\left[
3(\vec S_Q\!\cdot\!\vec n)(\vec S_{\bar Q}\!\cdot\! \vec n)-\vec
  S_Q\cdot\vec S_{\bar Q}\right],
\eea
where 
\bea
V_{SS}(r)=\frac{1}{3m_Q^2}\nabla^2 V_C(r),\,\,\,
V_{LS}(r)=\frac{1}{2m_Q^2}\frac1{r}\frac{dV_C(r)}{dr},
\,\,\,
V_{T}(r)=\frac{1}{3m_Q^2}\left(\frac{1}{r}\frac{dV_C(r)}{dr}
-\frac{d^2V_C(r)}{dr^2}\right).
\label{spin}\eea

\begin{figure}[h]
\centering
{\includegraphics[scale=0.4]{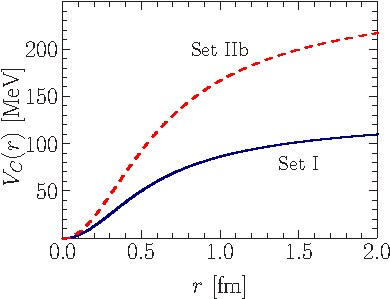}
\includegraphics[scale=0.4]{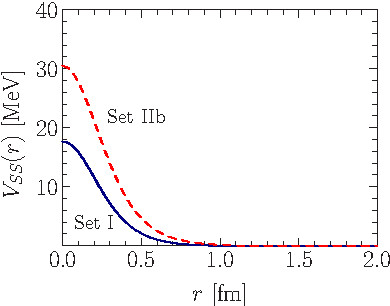}\\
\includegraphics[scale=0.4]{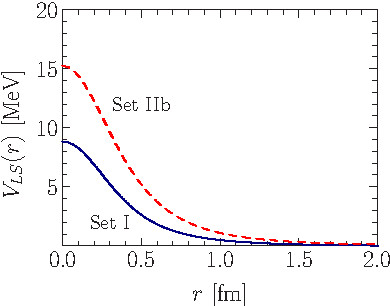}
\includegraphics[scale=0.4]{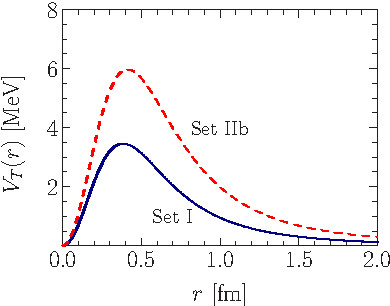}}
\caption{The components of  colorless state 
$c\bar c$ potential in ILM. Solid curve $\sim$ Set~I  
  $\bar\rho=0.33$~fm and ${\bar R}=1\,\mathrm{fm}$
 $\sim$  phenomenology~\cite{Diakonov:2002fq,Schafer:1996wv}.    
Dashed one $\sim$ Set~IIb  
  $\bar\rho=0.36$~fm, ${\bar R}=0.89\,\mathrm{fm}$ 
$\sim$ lattice~\cite{lattice} and ILM with $1/N_c$ corrections~\cite{Goeke:2007bj},
  $m_c=1275$~MeV.
} 
\label{potential}
\end{figure}

Fig.~\ref{potential} represent the result of numerical calculations of the components of  colorless state 
$c\bar c$ potential in ILM, corresponding to two sets
of the parameters -- average instanton size $\bar\rho$ and average inter-instanton distance $\bar R$.
\vskip -0.9cm
\begin{table}[h]
\centering
\begin{tabular}{c|c|c|c}
$\Delta M_{c\bar{c}}(J^P)$&Set~I& Set~IIb &  Exp.\\
\hline
$\Delta M_{\eta_c}(0^-)$ & 118,81 & 203,64 & $ 433,6\pm 0.6$\\
$\Delta M_{J/\psi}(1^-)$ & 119,57 & 205,36 & $546,916\pm 0.11$\\ 
$\Delta M_{\chi_{c0}}(0\dagger)$ &  142,43 & 250,86  & $864,75\pm 0.31$\\
\end{tabular}
\caption{ ILM contribution to the $(c\bar c)$ states. $\Delta M_{c\bar{c}}=M_{c\bar{c}}-2m_c$. }
\label{cbarc}
\end{table}

One can see from Table~\ref{cbarc} that the instanton effects at lowest states of $(c\bar c)$ quarkonium are not small $\sim 30-40\,\%$ in
comparison with the experimental data and strongly depend on instanton liquid parameters.

Since DLM pretend to describe QCD vacuum on the large distances too, it is natural to extend these calculations to the whole range of distances at the next step.

\section{Heavy and light quarks in ILM}

In the presence of light quarks the ILM heavy quark propagator is: 
\bea
&&S_H=\int\prod_{f}D\psi_{f}D\psi_{f}^{\dagger}\exp\int\left(\psi_{f}^{\dagger}(\hat{p}\,+\, im_{f})\psi_{f}\right)
\prod_\pm\left(\overline{\prod_{f}V_{\pm,f}[\psi^{\dagger},\psi]}\right)^{N_\pm} w[\psi,\psi^{\dagger}],
\\ \nonumber
&&w[\psi,\psi^{\dagger}]
=\left\{\prod_\pm\left(\overline{\prod_{f}V_{\pm,f}[\psi^{\dagger},\psi]}\right)^{N_\pm}\right\}^{-1}\int\prod_{\pm}^{N_{\pm}} d\zeta_\pm
\left\{\prod_{\pm}^{N_{\pm}}\prod_{f}V_{\pm,f}[\psi^{\dagger} ,\psi ]\right\}\frac{1}{\theta^{-1}-\sum_i a_i}.
\eea
Obvious extension of the Eq.~(\ref{pobylitsa}) is 
\bea
w^{-1}[\psi,\psi^{\dagger}]=\theta^{-1} + \sum_i\int d\zeta_i (w[\psi,\psi^{\dagger}]-a_i^{-1})^{-1}.
\label{pobylitsaextension}\eea
Again, in the lowest order on density we have
\bea
&&w^{-1}[\psi,\psi^{\dagger}]=\theta^{-1} + \frac{N}{2}\sum_\pm \frac{1}{\overline{\prod_{f}V_{\pm,f}[\psi^{\dagger},\psi]}}
\int d\zeta_\pm \prod_{f}V_{\pm,f}[\psi^{\dagger},\psi]\left( \theta-a_\pm^{-1}\right)^{-1}+ O(N^2/V^2)
\nonumber\\
&&=\theta^{-1} - \frac{N}{2}\sum_\pm \frac{1}{\overline{\prod_{f}V_{\pm,f}[\psi^{\dagger},\psi]}}\Delta_{H,\pm}[\psi^{\dagger},\psi ] + O(N^2/V^2),
\label{lowestextension}\eea
where 
\bea
\Delta_{H,\pm}[\psi^{\dagger},\psi ]=\int d\zeta_\pm\prod_{f}V_{\pm,f}[\psi^{\dagger},\psi]\theta^{-1}(w_\pm-\theta)\theta^{-1}
\eea
represent the interactions of single heavy and $N_f$ light quarks induced by
 (anti)instanton.  At $N_f=1$ we get the quark propagator in
the instanton media with account of light quarks as 
\bea
\label{SH1}
 S_H=\frac{1}{\theta^{-1} - \lambda\sum_\pm
   \Delta_{H,\pm}[\frac{\delta}{\delta\xi}
   ,\frac{\delta}{\delta\xi^\dagger}] } 
 \exp\left[-\xi^\dagger\left(\hat p \,+\, i(m+M(p))
   \right)^{-1}\xi\right]_{|_{\xi=\xi^\dagger=0}} .
\eea 
The heavy and $N_f$ light quark interaction term is  explicitly
 given by the
expression: 
\begin{eqnarray}
&&S_{Qq}= - \lambda\sum_\pm Q^{\dagger}\Delta_{H,\pm}[\psi^{\dagger},\psi ]Q=  
- i\lambda\sum_\pm\int d^4z_\pm dU_\pm \prod_{f=1}^{N_f}\frac{d^4 k_f}{(2\pi)^4}\frac{d^4 q_f}{(2\pi)^4}  \exp(i(q_f-k_f)z_\pm)
\nonumber\\
&&\times \frac{(2\pi\rho )^2 F(k_f )F(q_f )}{8}\psi_{f,a_f \alpha_f}^{\dagger}(k_f)
(\gamma_{\mu_f}\gamma_{\nu_f} \frac{1\pm\gamma_5}{2})_{\alpha_f \beta_f}
(U^{a_f}_{\pm,i_f}(\tau^{\mp}_{\mu_f}\tau^{\pm}_{\nu_f})^{i_f}_{j_f} U^{\dagger j_f}_{\pm,b_f} \psi^{b_f}_{f,\beta_f}(q_f)
\nonumber\\
\label{SQq}
&&\times Q_{a_3}^{\dagger} U^{a_3}_{\pm,i_3}\left(\theta^{-1}(w_\pm-\theta)\theta^{-1}\right)_{j_3}^{i_3}  U^{\dagger j_3}_{\pm,b_3}  Q^{b_3} .
\end{eqnarray}
Here matrix $U_\pm$ is a (anti)instanton color orientation.
It is evident that the integration over $z$ leads to the
energy-momentum conservation delta-function, while the integration
over color orientation provides the specific structure of the
interaction terms. Also, each light quark leg is accompanied by the
form-factor $F=F(k\rho)$, which is localized at the region $k\rho\le
1,$ as expected. 

\subsection{$Q\bar Q$ potential generated by light quarks exchange}
\begin{figure}[h]
\centering\includegraphics[scale=0.25]{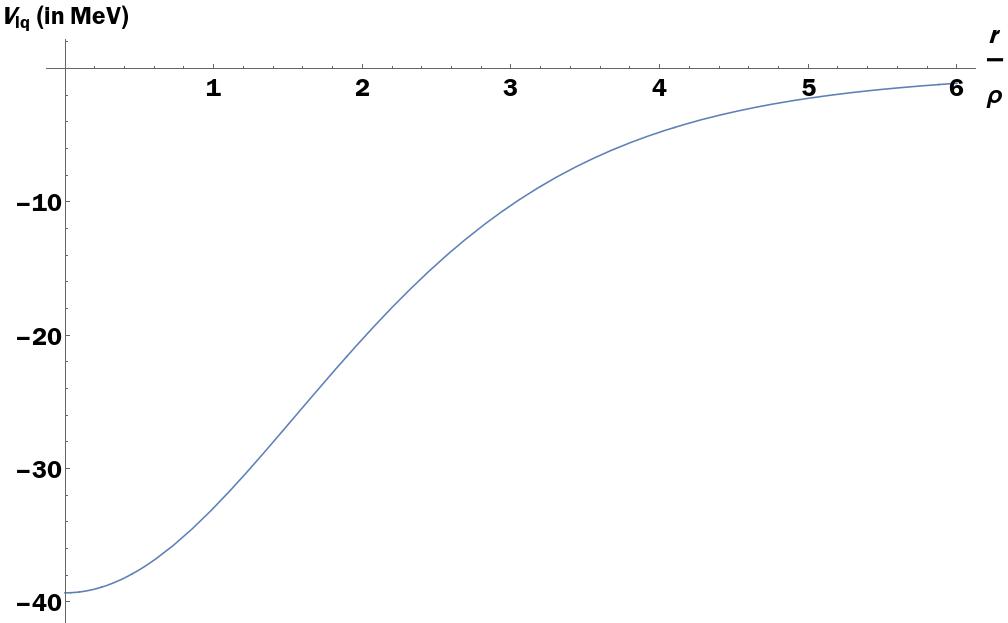}
\caption{ $Q\bar Q$ potential $V_{lq}(r/\rho)$ (in MeV) in singlet color state, generated by light quarks exchange ($N_f=1$). It is taken the parameters set~I  $\bar\rho=0.33$~fm, ${\bar R}=1\,\mathrm{fm}$ .}
\label{Vlq}\end{figure}

The consideration of the $Q\bar Q$ correlators (Wilson loops) in the same approach as given by
Eqs.(\ref{pobylitsaextension}-\ref{lowestextension}) provide the way for the calculations of $Q\bar Q$ potential $V_{lq}$ generated by light quarks exchange. The result of the calculations $V_{lq}(r)$ in simplest case $N_f=1$ and with the parameters set I is shown at Fig.~\ref{Vlq}.  

It is obvious that at $N_f=2$ case we have  to have two-pions exchange contribution to the  $V_{lq}(r)$.

\subsection{Heavy quark light mesons interaction term}

Heavy quark light quarks interaction term~(\ref{SQq}) has an essential part which is co-product of colorless heavy and light quarks factors. In the $N_f=2$ case it is
\bea
&&S[\psi,\psi^{\dagger},Q^{\dagger} Q] =   i\lambda\int d^4x \exp(-ipx)\frac{d^4
  p_1}{(2\pi)^4} \frac{d^4 p_2d}{(2\pi)^4}  
\frac{16 \pi \rho^3}{N_c}  i_0(p\rho) Q^\dagger( p_2) Q(p_1) 
\\\nonumber
&&\frac{1}{8(N_c^2-1)}\left[ \left(1-\frac{1}{2N_c}\right)\left(
(q^\dagger(x)q(x))^2 + (q^\dagger\gamma_5 q)^2-(q^\dagger\vec\tau q)^2
-(q^\dagger\gamma_5\vec\tau q)^2 \right)\right. 
\nonumber \\\nonumber
 &&\left.- \frac{1}{8N_c}\left((q^\dagger\sigma_{\mu\nu}q)^2 +(q^\dagger\gamma_5\sigma_{\mu\nu}q)^2 
 - (q^\dagger\sigma_{\mu\nu}\vec\tau q)^2 - (q^\dagger\gamma_5\sigma_{\mu\nu}\vec\tau q)^2\right) \right],
\eea
where $  i_0(y)=\int_0^\infty dz\, \frac{\sin[zy]}{zy} \left(z \cos\left[\frac{\pi z}{2\sqrt{(z^2 + 1)}}\right]\right)^2$, 
$q(x)=2\pi\rho F(i\partial )\psi(x) $ and  $q^\dagger(x)=2\pi\rho
F(i\partial )\psi^\dagger(x) $.
 
The application of the standard bosonization procedure to the light
quarks and the calculation of the path integrals over $\lambda$ and
meson fields in the saddle point approximation lead
to the vacuum equations in the leading order (LO) on $1/N_c$
expansion.  It is natural that these mesons $(\sigma,\vec\phi, ...)$
have the properties corresponding to the light
quarks bilinears $(q^{\dagger} q,q^{\dagger} \gamma_5\vec\tau q,...)$.   

So, the vacuum equations in the LO are
written as
\bea
\frac{1}{2}\Tr \frac{iM(p)}{\hat p+i(m+M(p))}=N=\frac{1}{2}\sigma^2_0 V, \,\,\, M(p) =
MF^2(p),\,\,\, M^2= (2\pi\rho
)^4\lambda\frac{2N_c-1}{2N_c(N_c^2-1)}\sigma^2_0. 
\label{VE}\eea
They fix the coupling $\lambda$ and the saddle-point $\sigma_0$ and
accordingly the dynamical quark mass $M(p)$. 

Now the total scalar meson field is $\sigma=\sigma_0+\sigma'$,
where $\sigma'$ is a quantum fluctuation. Other mesons are presented
only by their quantum fluctuations. 

At the saddle points we have the effective action for the mesons and
colorless heavy quark $Q^{\dagger} Q$ bilinear as 
\bea 
\nonumber
&&S[\sigma' ,\vec\phi' ,\eta' ,\vec\sigma',Q^{\dagger} Q]=
-\Tr \ln\frac{\hat p+i(m+M(p))}{\hat p+im} +N/2
+ \frac{1}{2}\int d^4
x\left({\sigma'}^{2}+{\vec{\phi'}}^2+{\vec{\sigma'}}^2+{\eta'}^2\right) 
\\\label{SQmesons}
&&
-\Tr\ln\left[1+\frac{1}{\hat
    p+i(m+M(p))}\frac{iM}{\sigma_0}F\left(\sigma'+i\gamma_5\vec\tau\vec\phi'+i\vec\tau\vec\sigma'+\gamma_5\eta'\right)F\right] 
\\\nonumber
&&+\Tr \frac{1}{\hat
  p+i(m+M(p))+\frac{iM}{\sigma_0}F\left(\sigma'+i\gamma_5\vec\tau\vec\phi'+i\vec\tau\vec\sigma'+\gamma_5\eta'\right)F} 
\\\nonumber
&&\times i\left(M(p)+\frac{M}{\sigma_0}F\left(\sigma'+i\gamma_5\vec\tau\vec\phi'+i\vec\tau\vec\sigma'+\gamma_5\eta'\right)F\right)
\left(\frac{i}{2} \Delta m_Q R^4
\int e^{-ipx}\frac{d^4 p_1}{(2\pi)^4} \frac{d^4 p_2}{(2\pi)^4}\frac{i_0( p\rho)}{i_0(0)}Q^\dagger Q\right) .
\eea
The first and second lines describe a mesons and their interactions, while the
third and forth one describe the renormalization of the heavy
quark mass and heavy quark-light quark mesons interactions terms.  

From the Eq.~(\ref{SQmesons}) we have the heavy quark-pion interaction term~\cite{M2014}:

\begin{eqnarray} 
S_{Q\pi}=i\Delta m_Q R^4\frac{F^2_{\pi Q}}{4}\int d^4x\,
\tr_f\partial_\mu U(x) \partial_\mu U^\dagger(x)
 \int e^{-ipx}\frac{d^4 p_1}{(2\pi)^4} \frac{d^4 p_2}{(2\pi)^4}
\frac{i_0( p\rho)}{i_0(0)} Q^\dagger ( p_2)
  Q( p_1)
\label{SQpions3}
\end{eqnarray}
where matrix pion field $U\approx(\sigma+i\vec\tau\vec\phi')/\sigma_0$,  $F^2_{\pi Q}\approx 0.7 F^2_{\pi}$ and $p=p_1-p_2.$

The similar approach for the calculations of $Q\bar Q$ correlators (Wilson loop) with account of light quarks will provide
the interaction term of the pair of heavy quarks with pions  $S_{QQ\pi}$. Both of these terms  $S_{QQ\pi}$ and  $S_{Q\pi}$ are responsible for the
 two-pions transitions in a heavy quarkoniums. We are planning to apply them in these transitions.

\section{Conclusion}

Lowest state hadrons naturally has a small size quark cores. 
In this case their properties insensitive  to the confinement and ILM is applicable.  
It was demonstrated by successful application of ILM to the light quark physics.
Main feature of light quarks is SBCS naturally described as would-be zero-mode dominance.
Light quarks strongly interact with instantons due to zero-modes. Almost massless quark become 
very massive with dynamical mass $M\sim 365$ MeV (at $\bar\rho=0.33$ fm, $\bar R=1$ fm), 
which can be considered as a strength of a light quark-instanton interaction.

On the other hand, heavy quarks much weaker interact with instantons. The analogous quantity -- 
the shift of a heavy quark mass $\Delta m_Q\sim 70$ MeV (at the same $\bar\rho,\bar R$) similarly 
can be considered as a strength of a heavy quark-instanton interaction.
Nevertheless the influence of the heavy quarks potential induced by instantons (see Fig.~\ref{QQpotential}) might be important for the spectra of heavy quarkonia as demonstrated  by Table~\ref{cbarc}.

It is natural that instantons generate also heavy and light quarks interactions with a strength $\sim M\Delta m_Q$. 
Such a vertexes leads to light quark exchange potential for heavy quarks (see Fig.~\ref{Vlq}). 
The light-heavy quarks interactions terms naturally leads to the light hadrons transitions in a heavy quarkonia and to the light-heavy quarks bound states -- mesons and baryons.

There is a list of a problems which can be solved within this approach.

\end{document}